\documentclass[a4paper, oneside, twocolumn, notitlepage, 10pt]{extarticle_ecoc}
\usepackage{ecoc}
\usepackage{subcaption}
\renewcommand{\supercite}[1]{\textsuperscript{\cite{#1}}}

\begin{document}
\selectlanguage{english}    


\title{DPD-KAN: Kolmogorov-Arnold Networks for Low Complexity Digital Predistortion in 5G Analog Radio-over-Fiber Systems}%

\author{
    Bilal Khalid\textsuperscript{(1)}, Fabio Cavaliere\textsuperscript{(2)}, Luca Giorgi\textsuperscript{(2)}, Pedro Freire \textsuperscript{(1)}, Sergei K. Turitsyn\textsuperscript{(1)} \\ and Jaroslaw E. Prilepsky\textsuperscript{(1)}
}

\maketitle                  


\begin{strip}
    \begin{author_descr}

        \textsuperscript{(1)} Aston Institute of Photonic Technologies, Aston University, Birmingham, UK,
        \textcolor{blue}{\uline{r.khalid4@aston.ac.uk}}

        \textsuperscript{(2)} Ericsson Research, Ericsson, 56124, Pisa, Italy
\vspace{-2mm}
    \end{author_descr}
\end{strip}

\renewcommand\footnotemark{}
\renewcommand\footnoterule{}


\begin{strip}
    \begin{ecoc_abstract}
        We demonstrate the first KAN-based DPD model for 5G analog RoF fronthaul link, achieving a $24.2\%$ lower EVM than multi-layer perceptron and $29.6\%$ lower than Volterra-based GMP at equivalent Bit Operations. To attain an EVM below $2\%$, KAN requires $\approx52\%$ fewer BOPs than a perceptron. ©2026 The Author(s) 
        \vspace{-2mm}
    \end{ecoc_abstract}
\end{strip}


\begin{figure*}[!b]
  \centering
  \vspace{-2mm}
  \includegraphics[width=1.02\textwidth]{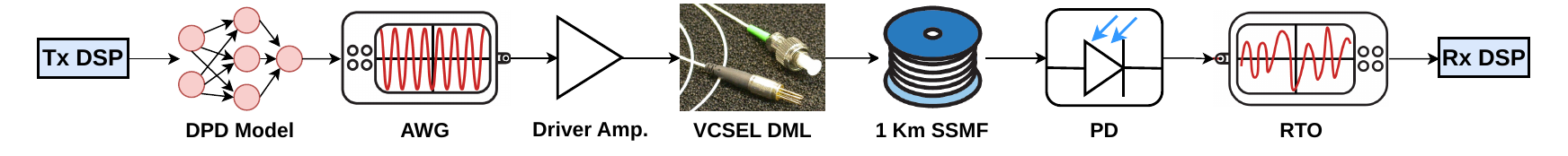}
  \caption{System model for VCSEL-based Analog RoF link.}
  \label{fig:system}
\end{figure*}

\section{Introduction}
Radio-over-fiber (RoF) has emerged as a key enabling technology for the distribution of wireless signals in dense 5G and prospective 6G network architectures\supercite{lim2025past}. While Digital RoF (D-RoF) guarantees high signal fidelity, it incurs a significant bandwidth overhead and faces the challenges of increased power consumption and heat dissipation for dense Remote Radio Units (RRU) deployment scenarios\supercite{sabella2025microwave}. Analog RoF (A-RoF) is the alternative option that offers high bandwidth efficiency, lower latency, and reduced complexity at the RRU\supercite{thomas2015performance}. However, A-RoF links are highly susceptible to signal degradation due to several impairments. In A-RoF systems utilizing directly modulated lasers (DMLs), these impairments primarily originate from laser saturation and clipping, intensity-dependent frequency chirp, and the subsequent chirp-dispersion interaction within the standard single-mode fiber (SSMF)\supercite{thomas2015performance}. 

Digital predistortion (DPD) is one of the techniques commonly used to mitigate these effects. Machine learning (ML)-based architectures have heavily influenced recent DPD research. ML-based models\supercite{hadi2019digital, hadi2021neural} have been shown to outperform classical DPD methods based on Volterra series, such as Memory Polynomials (MP)\supercite{ding2004mp1,atso2012mp2} and Generalized Memory Polynomials (GMP)\supercite{morgan2006gmp1}. Techniques utilizing Multilayer Perceptrons (MLPs)\supercite{pereira2022machine, najarro2018nonlinear,silva2025performance}, Convolutional Neural Networks (CNNs)\supercite{hadi2022cnn}, and Recurrent Neural Networks (RNNs)\supercite{pereira2023amplified, hadi2024digital} have been developed to achieve effective linearization. However, the power consumption and computational complexity of ML models remain a concern for practical deployment in fronthaul RoF links\supercite{sabella2025microwave, he2020machine, silva2025performance}.

Kolmogorov-Arnold Networks (KANs)\supercite{liu2025kan} have recently emerged as an alternative to traditional neural networks. Unlike MLPs, which use fixed activation functions at the nodes, KANs place learnable activation functions on the edges. This allows KANs to learn non-linear relationships while maintaining greater interpretability\supercite{somvanshi2025survey}. Recent works demonstrate the utility of KANs in baseband equalization for purely optical IM-DD systems\supercite{chen2025neural} and DPD for standalone radio frequency (RF) power amplifiers\supercite{jinghao2025real}. However, to the best of our knowledge, KANs have not previously been investigated for DPD in A-RoF systems, which presents a fundamentally different challenge.

In this paper, we demonstrate the first KAN-based DPD model (DPD-KAN) for a 5G NR A-RoF fronthaul link. We evaluate the proposed method against MLP and conventional GMP baselines. The computational complexity is measured in terms of Bit Operations (BOPs), which accurately capture the complexity of quantized networks\supercite{freire2024computational, khalid2026hardware}. Results indicate that KANs achieve superior performance as compared to the baselines in terms of error vector magnitude (EVM), adjacent channel leakage ratio (ACLR) and normalized mean square error (NMSE). Importantly, DPD-KAN converges to the optimal performance at $52\%$ lower BOPs than MLP, highlighting its suitability and benefits for complexity-constrained real-world deployment.

\begin{figure}[!t]
   \centering
   \includegraphics[width=\columnwidth]{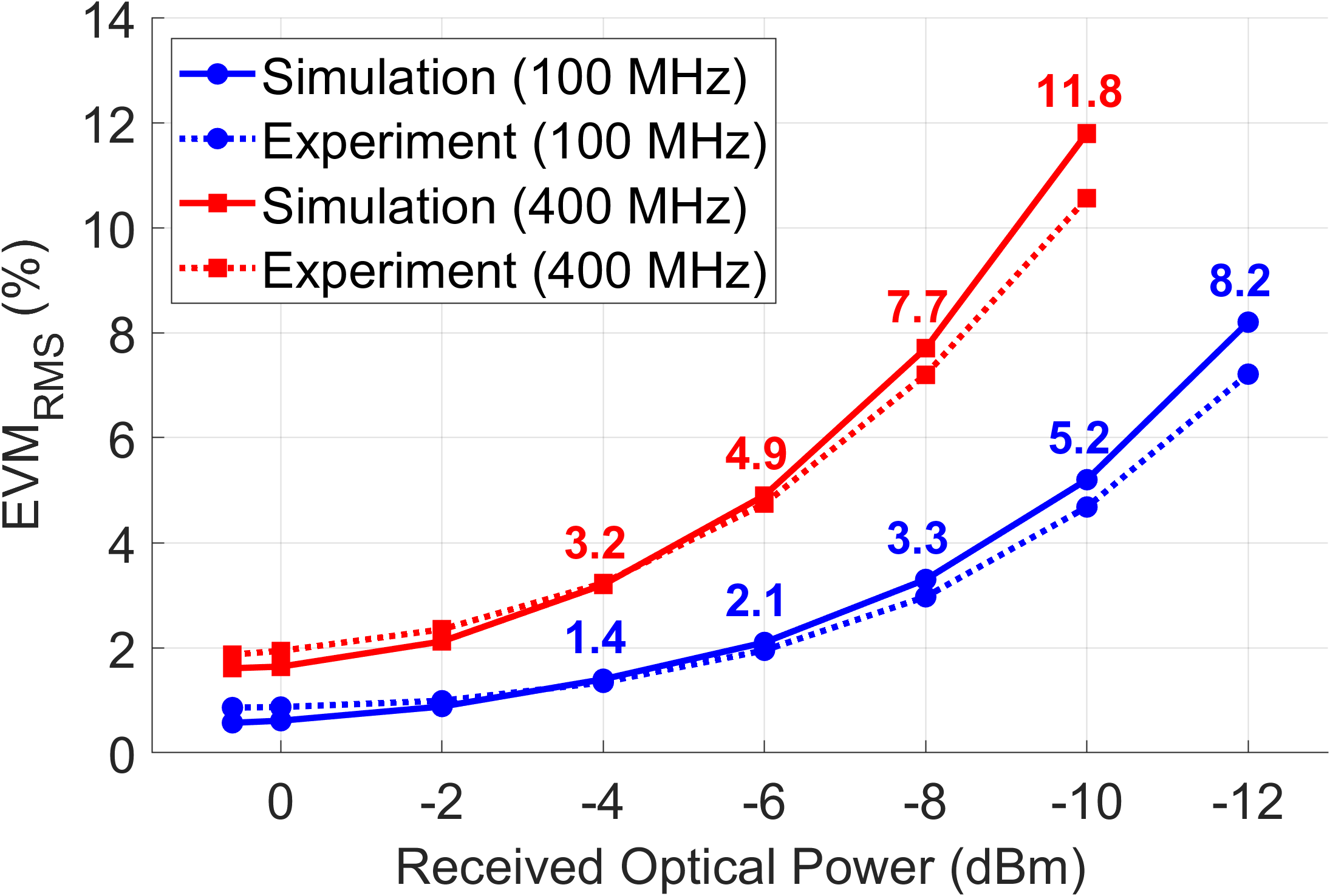}
   \caption{Comparison of simulation and experimental results.}
   \label{fig:model_verification}
\end{figure}

\begin{figure}[!t]
\vspace{5pt}
   \centering
   \includegraphics[width=\columnwidth]{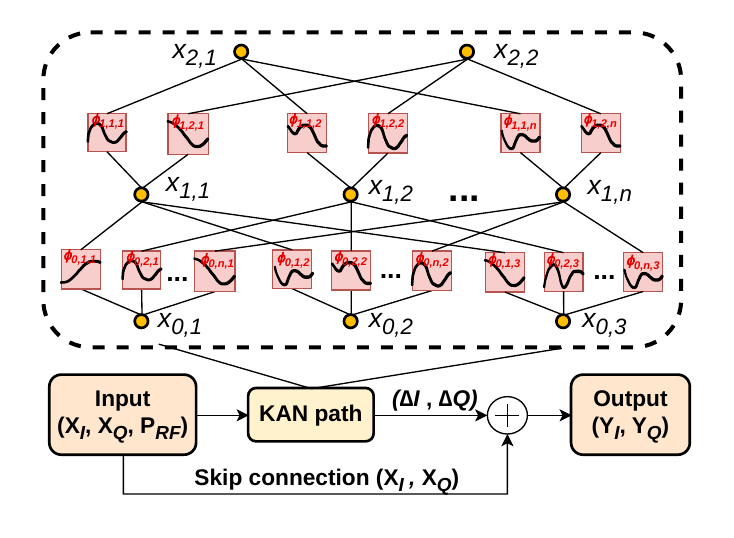}
   \caption{DPD-KAN architecture.}
   \label{fig:dpd_kan}
\end{figure}

\section{Experimental Setup}

The system model, illustrated in Fig.~\ref{fig:system}, realizes a 5G NR A-RoF fronthaul link. We consider a 100\,MHz bandwidth, 64-QAM, 5G NR baseband waveform conforming to 3GPP Test Model (TM) 3.1 that is upconverted to an RF carrier frequency of 3.6\,GHz using an arbitrary waveform generator (AWG). A driver amplifier is used to amplify the signal before it reaches the optical transmitter. The optical source is a VERTILAS VL-1550-25G-SH-H4 VCSEL directly modulated laser (DML) with a wavelength of 1550\,nm. We consider a fiber length of $1$\,km as more than $97\%$ of deployed fronthaul links fall within this operational range\supercite{mopa2025v26a, cavaliere2025centralization}. At the receiver, the signal is captured by a real-time oscilloscope (RTO) after photo-detection.

We developed the link simulation model in MATLAB. Prior to generating the training dataset, the MATLAB model was validated against experimental measurements. Figure~\ref{fig:model_verification} shows the EVM versus received optical power plot for the described transmission setup for $100$~MHz and $400$~MHz signals. The close agreement between the simulation and experimental results confirms the correctness of the simulation model. To train the DPD models, Indirect Learning Architecture (ILA)\supercite{eun1997ila} is used.

\section{DPD-KAN Architecture}
The KAN DPD model shown in Fig.~\ref{fig:dpd_kan} processes a three-dimensional input vector consisting of the In-phase (I) and Quadrature (Q) components of the complex baseband signal, concatenated with the RF input power. It employs a dual-path residual architecture to improve training stability and convergence. The linear path is an identity mapping that directly forwards the uncorrected signal. The non-linear KAN path is tasked with learning the residual correction $(\Delta I, \Delta Q)$. Its learnable activation functions $\phi(x)$ are represented as
\vspace{-2mm}
\begin{equation}
\label{eq:kan_edge_bspline}
\phi(x) = w_b \, \text{SiLU}(x) + \sum_{i=0}^{G+k-1} \! \! w_i \, B_{i,k}(x) \, ,
\end{equation}
where $w_b$ is the base weight, $w_i$ denotes the basis control coefficients, $k$ is the spline order, $G$ represents the grid size and $B_{i,k}(x)$ are the B-spline basis functions. We employ a memoryless architecture because memory effects induced by chromatic dispersion are negligible for $1$\,km link length. Consequently, the dominant system impairments are static, namely the electro-optic conversion nonlinearity of the VCSEL and potential saturation of the driver amplifier. We verified this by adding memory taps that yielded no significant performance improvement. To maintain consistency, MLP also employs the same residual learning architecture. The KAN model with a complexity of $\approx10^4$ BOPs has a single hidden layer with $n=7$ nodes, spline order $k=3$ and grid size $G=7$. The trained model weights are uniformly quantized to an $8$-bit resolution during inference. 

\begin{figure}[!t]
   \centering
   \includegraphics[width=\columnwidth]{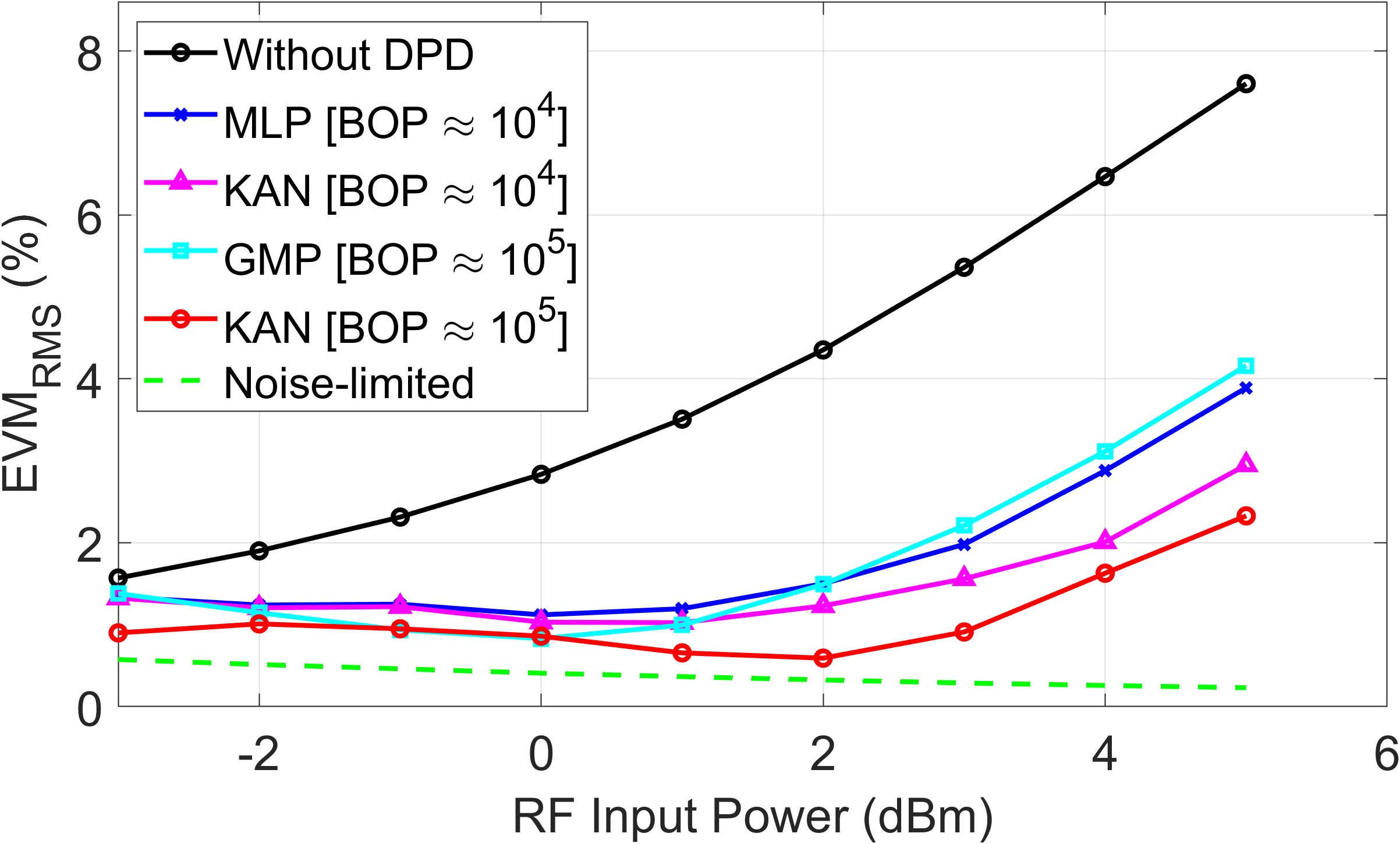}
   \caption{EVM\textsubscript{RMS} vs. RF input power. DPD-KAN gives the best results at $10^4$ BOPs, with further improvement at $10^5$ BOPs.}
   \label{fig:evm}
\end{figure}

\begin{figure}[!t]
   \vspace{10pt}
   \centering
   \includegraphics[width=\columnwidth]{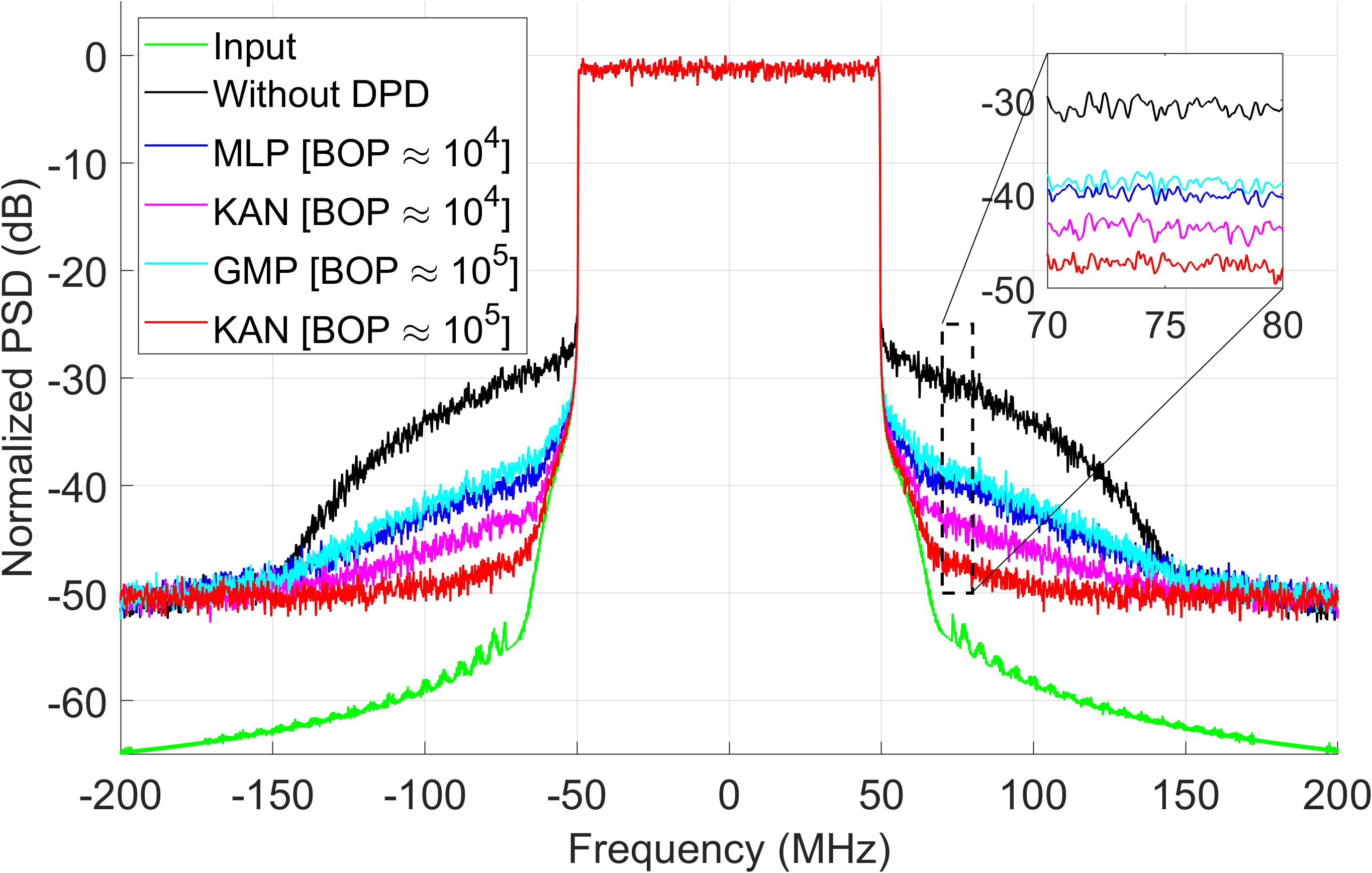}
   \caption{PSD plot at $3$\,dBm RF input power showing $\approx\!5$\,dB advantage of DPD-KAN at low complexity.}
   \label{fig:psd}
\end{figure}

\vspace{-2mm}
\section{Results and Discussion}
We compare the linearization performance of KAN with MLP and the conventional GMP methods, and present a comprehensive performance-versus-complexity analysis.
\vspace{3pt}
\\\textit{Linearization performance.} Figure~\ref{fig:evm} shows the effect on EVM when the RF signal input power is varied from $-3$ to $+5$\,dBm. Without DPD, EVM rises rapidly from $1.6\%$ to $7.6\%$, which reflects increased nonlinear distortion at higher RF signal powers. We observe that at low complexity of $\approx10^4$ BOPs, KAN outperforms both MLP and GMP. The gap widens at higher input powers where the stronger nonlinearity demands greater model expressivity. At $5$\,dBm, KAN achieves a $24.2\%$ lower EVM  than an equivalent complexity MLP and $29.6\%$ lower than a GMP DPD model with  a higher complexity of $10^5$ BOPs. This demonstrates that KAN's learnable B-spline activations provide a more efficient nonlinear basis for the distortion profile of the VCSEL-based A-RoF link than fixed-activation MLP and Volterra-based GMP under stringent complexity constraints. At a higher complexity of $10^5$ BOPs, KAN's performance further improves, maintaining a low EVM across the full power range. At this level of complexity, MLP also achieves similar performance. This is illustrated in Table~\ref{tab:performance_3dbm}, which summarizes the EVM, ACLR and NMSE at $3$\,dBm. At BOP\,$\approx\!10^{5}$, KAN achieves an EVM of $0.91$\%, an ACLR of $-47.46$~dBc, and an NMSE of $-40.81$\ dB, which is almost indistinguishable from an equivalent complexity MLP. However, KAN converges to the optimal performance at significantly lower complexity than MLP, as discussed in the following paragraphs. 

\begin{figure*}[!t]
    \centering
    \subfloat[\label{fig:pareto}]{%
        \includegraphics[width=0.48\textwidth]{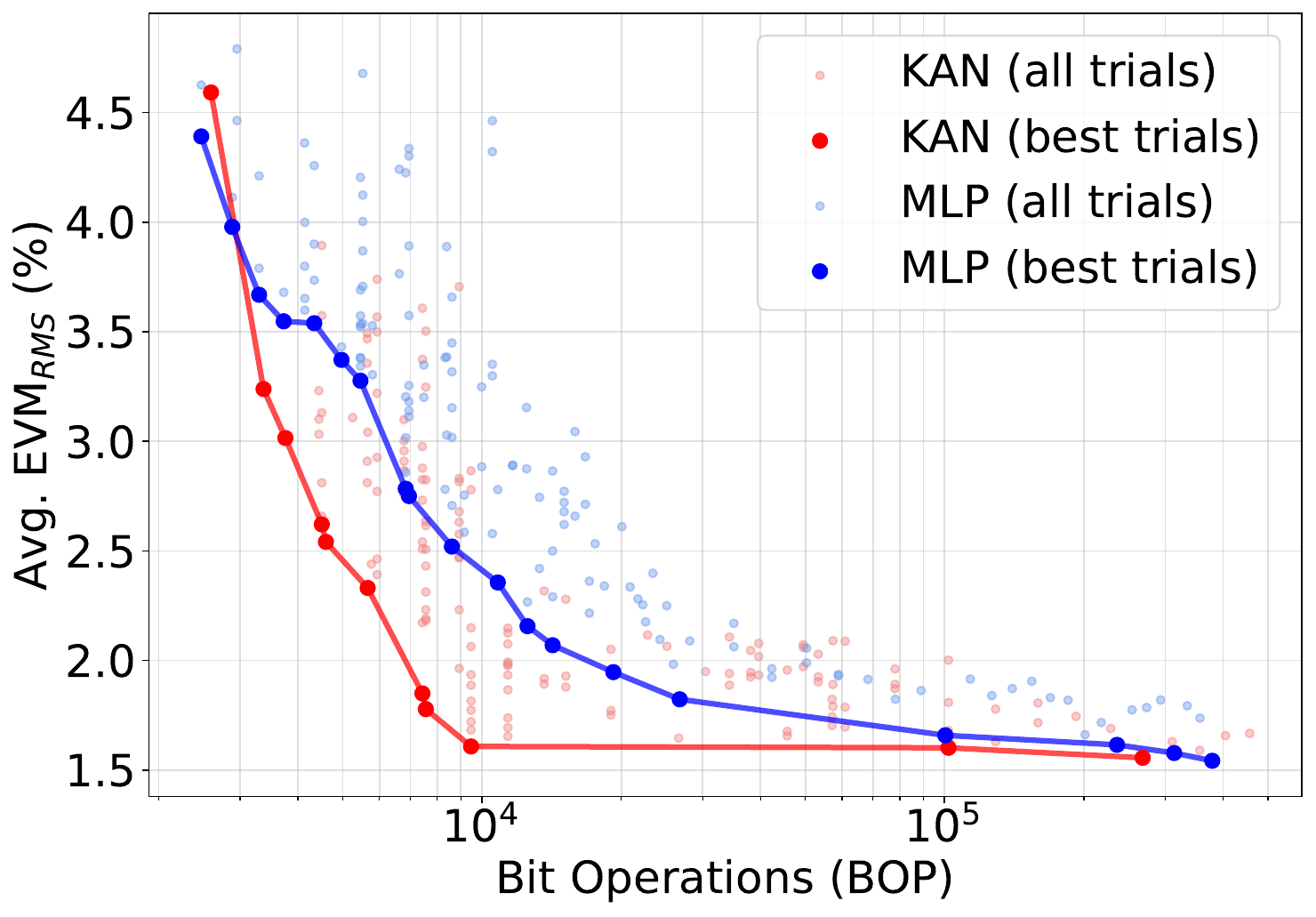}
    }
    \hfill
    \subfloat[\label{fig:average}]{%
        \includegraphics[width=0.48\textwidth]{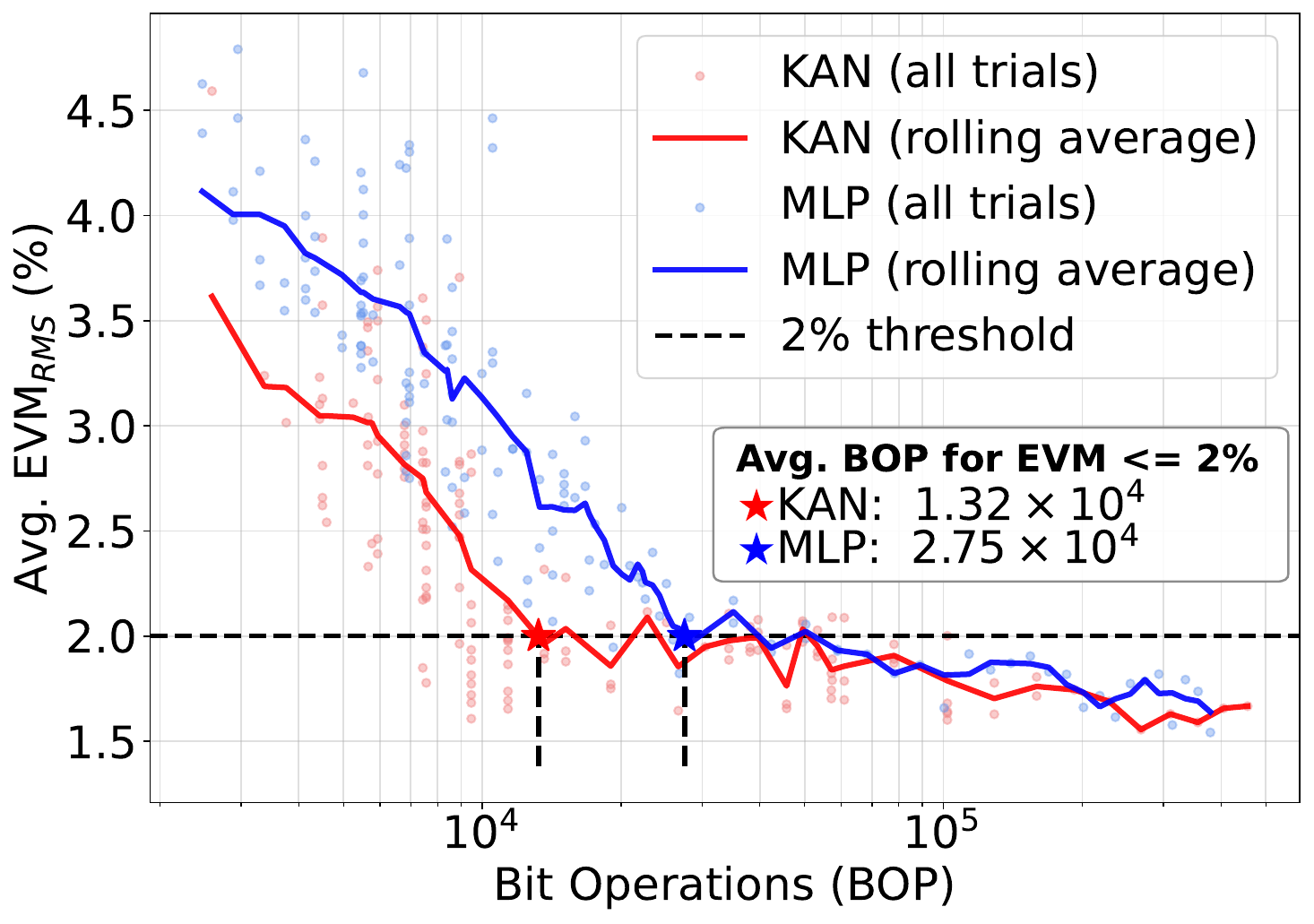}
    }
    \vspace{1em}
    \caption{Performance vs. complexity trade-off for KAN and MLP across 200 independent trials. (a) Best performing trials (b) Rolling average of EVM across the complexity interval. KAN reaches $2\%$ EVM threshold consuming $52\%$ fewer BOPs as compared to MLP.}
    \label{fig:combined_complexity}
\end{figure*}

\begin{table}[htbp]\vspace{-2mm}
\caption{Performance Comparison at 3 dBm RF Power}
\label{tab:performance_3dbm}
\centering
\setlength{\tabcolsep}{4pt} 
\begin{tabular}{|l|c|c|c|}
\hline
\textbf{DPD} & \textbf{EVM (\%)} & \textbf{ACLR (dBc)} & \textbf{NMSE (dB)} \\
\hline
--- & 5.35 & -34.54 & -25.42 \\
\hline
\multicolumn{4}{|c|}{BOP $\approx 1\times10^4$} \\
\hline
MLP & 1.98 & -42.47 & -34.06 \\
\textbf{KAN} & \textbf{1.56} & \textbf{-45.21} & \textbf{-36.14} \\
\hline
\multicolumn{4}{|c|}{BOP $\approx 1\times10^5$} \\
\hline
GMP & 2.21 & -41.57 & -33.09 \\
MLP & 0.93 & -47.07 & -40.63 \\
\textbf{KAN} & \textbf{0.91} & \textbf{-47.46} & \textbf{-40.81} \\
\hline
\end{tabular}
\end{table}

Figure~\ref{fig:psd} shows the normalized power spectral density (PSD) at an RF input power of $3$\,dBm. The PSD plot indicates high spectral regrowth without DPD. A $10^4$ BOPs KAN achieves approximately $5$\,dB reduced spectral regrowth as compared to a $10^4$ BOPs MLP and $10^5$ BOPs GMP. The KAN model at higher complexity of $10^5$ BOPs gives the best performance as highlighted in the inset of Fig.~\ref{fig:psd}. Thus, these results illustrate that KANs are suitable for efficient linearization of A-RoF links. 
\vspace{3pt}
\\\textit{Performance-versus-complexity analysis of KAN and MLP.} To compare the performance of KAN and MLP across different complexity levels, we performed $200$ independent trials for each architecture using the hyperparameter search space summarized in Table~\ref{tab:hyperparameters}. For each trial, we calculate the EVM for RF powers from $2-5$\,dBm and report the average EVM with respect to complexity in Fig.~\ref{fig:combined_complexity}. Fig.~\ref{fig:pareto} shows the best-performing trials for both networks. It can be seen that KAN is able to achieve much better performance at lower complexity. To better illustrate this, Fig.~\ref{fig:average} plots the rolling average of EVM across the complexity range. The results indicate that KAN converges at a significantly lower BOP. To reach an EVM threshold of less than $2\%$, KAN requires $52\%$ fewer BOPs than MLP. At high complexity, both KAN and MLP converge to a similar performance level.

\begin{table}[htbp]\vspace{-2mm}
\caption{Summary of hyperparameter search space}
\label{tab:hyperparameters}
\centering
\setlength{\tabcolsep}{4pt} 
\begin{tabular}{|l|c|c|}
\hline
\textbf{Hyperparameter} & \textbf{MLP} & \textbf{KAN} \\
\hline
Number of trials & $200$ & $200$ \\
Batch size & $1024$ & $1024$ \\
Learning rate & $1\times10^{-3}$ & $1\times10^{-3}$ \\
Hidden layers & $1, 2$ & $1, 2$ \\
Nodes per layer & $4$--$64$ & $4$--$32$ \\
Quantization bits & $6, 8$ & $6, 8$ \\
Training epochs & $40$ & $40$ \\
Activation & \footnotesize ReLU, GELU, SiLU & \footnotesize Learnable \\
Spline order & -- & $2, 3$ \\
Grid size & -- & $5, 7, 9$ \\
\hline
\end{tabular}
\end{table}

\vspace{-4mm}
\section{Conclusion}
We demonstrate linearization of a 5G NR A-RoF fronthaul link using KAN-based DPD. Results indicate that KAN outperforms conventional MLP and Volterra-based GMP models at low complexity. Specifically, to reach an EVM threshold below $2\%$, KAN requires $52\%$ fewer BOPs than MLP. Our results highlight that KANs offer an efficient architecture for low-complexity linearization of fronthaul A-RoF links.
\clearpage
\section{Acknowledgements}
This research is funded by EU's Horizon Europe research and innovation programme MSCA-DN NESTOR (G.A. 101119983). S. K. Turitsyn acknowledges EPSRC project TRANSNET (EP/R035342/1). Experiments were run on EPS ML Server (Grant EP/V036106/1). The authors have applied a Creative Commons Attribution (CC BY) licence to any Author Accepted Manuscript (AAM) version arising from this submission.


\printbibliography

\vspace{-4mm}

\end{document}